\documentclass[twocolumn,aps,prd,amsfonts,amsmath,amssymb,nofootinbib,longbibliography,preprintnumbers]{revtex4-1}

\usepackage{wasysym}
\usepackage{textcomp}
\usepackage{xcolor, cancel, ulem}
\usepackage{graphicx}
\usepackage{hyperref}
\usepackage{pdfpages}

\def\frac#1#2{{{{#1}}\over{{#2}}}}

\newsavebox{\ns}
\newsavebox{\dbrane}
\newsavebox{\dbshort}

\usepackage{epsfig}
\usepackage{amssymb}

\def\appendix{{\newpage\section*{Appendix}}\let\appendix\section%
        {\setcounter{section}{0}
        \gdef\thesection{\Alph{section}}}\section}

\newcommand\ba{\begin{eqnarray}}
\newcommand\ea{\end{eqnarray}}

\definecolor{DarkGreen}{rgb}{0,.64,0}
\definecolor{gunmetal}{rgb}{0.171875, 0.207031, 0.222656}
\definecolor{chartreuse}{rgb}{.49,.98,0}
\definecolor{amethyst}{rgb}{0.59375,0.398438,0.792969}
\definecolor{brownrust}{rgb}{0.6875, 0.316406, 0.242188}
\definecolor{Violet}{rgb}{0.5,0,1}
\definecolor{BurntOrange}{rgb}{0.792969,0.332031,0}
\definecolor{FreshEggplant}{rgb}{0.59375, 0., 0.414063}
\definecolor{salmon}{rgb}{0.996094,0.507813,0.410156}
 \definecolor{FrenchRose}{rgb}{0.96875, 0.292969, 0.5625}
\definecolor{Cabaret}{rgb}{0.808594, 0.242188, 0.46875}
\definecolor{Shamrock}{rgb}{0.242188, 0.808594, 0.582031}
\definecolor{RobinsEggBlue}{rgb}{0., 0.792969, 0.792969}
\definecolor{GuardsmanRed}{rgb}{0.792969, 0., 0.}
\definecolor{Sapphire}{rgb}{0.183594, 0.328125, 0.621094}
\definecolor{Sorbus}{rgb}{0.996094, 0.429688, 0.0273438}
\definecolor{Red}{rgb}{1,0,0}
\definecolor{Blue}{rgb}{0,0,1}
\definecolor{Green}{rgb}{0,1,0}
\definecolor{thistle3}{rgb}{0.800781, 0.707031, 0.800781}
\definecolor{thistle4}{rgb}{0.542969, 0.480469, 0.542969}

\newcommand{\nn}{\nonumber}

\def\Dslash{\,\,{\raise.15ex\hbox{/}\mkern-12mu D}}
\def\Dbarslash{\,\,{\raise.15ex\hbox{/}\mkern-12mu {\bar D}}}
\def\delslash{\,\,{\raise.15ex\hbox{/}\mkern-9mu \partial}}
\def\delbarslash{\,\,{\raise.15ex\hbox{/}\mkern-9mu {\bar\partial}}}
\def\pslash{\,\,{\raise.15ex\hbox{/}\mkern-9mu p}}
\def\calDslash{\,\,{\raise.15ex\hbox{/}\mkern-12mu {\cal D}}}

\newcommand{\hh}{{1\over 2}}

\renewcommand{\ll}{_}
\newcommand{\uu}{^}
\newcommand{\pp}{\partial}

\renewcommand{\d}{\delta}

\renewcommand{\dag}{^\dagger}

\newcommand{\s}{\sigma}

\renewcommand{\a}{\alpha}

\newcommand{\sqd}{^2}
\newcommand{\zb}{{\bar{z}}}

\renewcommand{\hh}{{1\over 2}}

\newcommand{\eee}[1]{\ba{#1}\ea}

\newcommand{\llsk}{\hskip .5in}

\newcommand{\pr}{^\prime}

\newcommand{\IZ}{\relax\ifmmode\mathchoice
{\hbox{\cmss Z\kern-.4em Z}}{\hbox{\cmss Z\kern-.4em Z}}
{\lower.9pt\hbox{\cmsss Z\kern-.4em Z}} {\lower1.2pt\hbox{\cmsss
Z\kern-.4em Z}}\else{\cmss Z\kern-.4em Z}\fi} \font\cmss=cmss10
\font\cmsss=cmss10 at 7pt
\newcommand{\inbar}{\,\vrule height1.5ex width.4pt depth0pt}
\newcommand{\IC}{{\relax\hbox{$\inbar\kern-.3em{\rm C}$}}}
\newcommand{\IQ}{{\relax\hbox{$\inbar\kern-.3em{\rm Q}$}}}
\newcommand{\IP}{\relax{\rm I\kern-.18em P}}
\def\icp#1{\IC\IP({#1})}

\newcommand{\ed}{\dot{e}}

\newcommand{\cc}{{\cal C}}

\renewcommand{\cc}{{c_1}}

\renewcommand{\cc}{c}

\renewcommand{\pr}{^\prime}

\newcommand{\IR}{\relax{\rm I\kern-.18em R}}
\def\blfootnote{\xdef\@thefnmark{}\@footnotetext}

\renewcommand{\cc}[1]{\cite{#1}}

\renewcommand{\bm}{\begin{matrix}}
\renewcommand{\em}{\end{matrix}}

\newcommand{\ee}[1]{\ba {#1} \ea}

\newcommand{\rr}[1]{(\ref{{#1}})}
\newcommand{\bbb}{\ba\begin{array}{c}}
\renewcommand{\eee}{\nonumber\end{array}\ea}
\newcommand{\een}[1]{\label{#1}\end{array}\ea}

\def\bi{\begin{itemize}}
\def\ei{\end{itemize}}

\def\ed{\end{document}}

\def\cc{{\cal C}}

\renewcommand{\rr}[1]{(\ref{#1})}

\def\cc{\,}

\def\cc{\,}

\usepackage{fancyhdr}
\renewcommand{\eqref}[1]{\rr{#1}}
\usepackage{dsfont}

\renewcommand{\bm}{\begin{matrix}}
\renewcommand{\em}{\end{matrix}}

\newcommand{\aaa}[1]{}

\def\be{\begin{eqnarray}}
\def\ee{\end{eqnarray}}
\def\nn{\nonumber}

\newcommand{\WhiteOutWithNotification}[1]{ {  \color{Red}  (  A SECTION HAS BEEN WHITED OUT HERE. \cc   )  } }
\newcommand{\WhiteOut}[1]{}

\def\kko{\ ,}
\def\ppo{\ .}

\def\redlowdash{{\color{Red}{\rule[-0.5ex]{2pt}{0.4pt}}}}
\def\redmiddash{{\color{Red}{\rule[+0.5ex]{2pt}{0.4pt}}}}

\def\cute{{\lower3.5pt\hbox{\sixly
  \kern-.21pt \char58 \kern-.21pt }}}

\def\midcute{{\lower-1.0pt\hbox{\sixly
  \kern-.21pt \char58 \kern-.21pt }}}

  \def\lowcute{{\lower3.5pt\hbox{\sixly
  \kern-.21pt \char58 \kern-.21pt }}}
  
  \def\redmidcute{{\color{Red} \midcute}}
  
  \def\redlowcute{{\color{Red} \lowcute}}
   
    \def\bluelowcute{{\color{Blue} \lowcute}}

   \def\swave{\bgroup \markoverwith \midcute \ULon} 
  
  \def\redswave{\bgroup \markoverwith \redmidcute \ULon} 
  
  \def\reduline{\bgroup \markoverwith \redlowdash \ULon}
  
   \def\blueuline{\bgroup \markoverwith \bluelowdash \ULon}
  
   \def\reduwave{\bgroup \markoverwith \redlowcute \ULon}
   
   \def\blueuwave{\bgroup \markoverwith \bluelowcute \ULon}
  
  \def\redsout{\bgroup \markoverwith \redmiddash \ULon}
  
   \def\bluesout{\bgroup \markoverwith \bluemiddash \ULon}

\let\oldcancel\cancel
\renewcommand\cancel[1][black]{%
  \def\CancelColor{\color{#1}}%
  \oldcancel}
  
  \let\oldxcancel\xcancel
\renewcommand\xcancel[1][black]{%
  \def\CancelColor{\color{#1}}%
  \oldxcancel}
  
   \let\oldbcancel\bcancel
\renewcommand\bcancel[1][black]{%
  \def\CancelColor{\color{#1}}%
  \oldbcancel}

\def\aat{\circ}

\def\hol{^{({\rm hol.})}}
\def\ant{^{({\rm ant.})}}

\begin{document}

\preprint{IPMU13-0170}

\title{Charge Quantization\\in the\\\hspace{.0in} \icp 1 Nonlinear $\s$-Model}

\author{Simeon Hellerman}
\email{simeon.hellerman.1@gmail.com}
\author{John Kehayias}
\email{john.kehayias@ipmu.jp}
\author{Tsutomu T.~Yanagida}
\email{tsutomu.tyanagida@ipmu.jp}

\affiliation{Kavli Institute for the Physics and Mathematics of the Universe (WPI)\\
Todai Institutes for Advanced Study, The University of Tokyo\\
Kashiwa, Chiba  277-8582, Japan}

\begin{abstract}

\noindent
We investigate the consistency conditions for matter fields coupled to
the four-dimensional (${\cal N} = 1$ supersymmetric) \icp 1 nonlinear
sigma model (the coset space $SU(2)_G/U(1)_H$). We find that
consistency requires that the $U(1)_H$ charge of the matter be
quantized, in units of half of the $U(1)_H$ charge of the
Nambu-Goldstone (NG) boson, if the matter has a nonsingular kinetic
term and the dynamics respect the full group $SU(2)_G$.  We can then
take the linearly realized group $U(1)_H$ to comprise the weak
hypercharge group $U(1)_Y$ of the Standard Model. Thus we have charge
quantization without a Grand Unified Theory (GUT), completely avoiding
problems like proton decay, doublet-triplet splitting, and magnetic
monopoles. We briefly investigate the phenomenological implications of
this model-building framework. The NG boson is fractionally charged
and completely stable. It can be naturally light, avoiding constraints
while being a component of dark matter or having applications in
nuclear physics. We also comment on the extension to other NLSMs on
coset spaces, which will be explored more fully in a followup paper.

\end{abstract}

\maketitle

\section{Introduction}

The Standard Model (SM) is an extremely successful theory of particle
physics, which has just seen a strong confirmation in the discovery of
the Higgs boson. However, the SM has many seemingly arbitrary
parameters and unexplained aspects. It is clear that some extension(s)
to the SM is necessary to explain such phenomena as neutrino masses,
dark matter, dark energy, quantum gravity and so on. That being said,
the success of the SM means that it must be a low energy limit of any
more complete theory which is valid at higher energies.

One of the most obvious (and earliest) frameworks for a completion of
the SM is a grand unified theory (GUT), with or without supersymmetry
(SUSY). Grand unified theories, if correct, would explain many
otherwise mysterious aspects in the matter representations chosen by
nature. In particular, they would explain the quantization of weak
hypercharge $U(1)\ll Y$, which translates into a quantization of the
electromagnetic charge $U(1)\ll{em}$ once the electroweak group $SU(2)
\times U(1)\ll Y$ is broken down to $U(1)\ll{em}$. This point was made
in the original proposal of GUTs by Georgi and Glashow
\cite{Georgi:1974sy}. This is an elegant answer to the longstanding
question of charge quantization, which comes naturally on the road to
a unified theory of all fundamental particles and forces.

However, the GUT framework has many seemingly intrinsic problems.  GUT
models tend to predict phenomena, such as cosmic monopole production
and baryon decay, which are observationally excluded at the order of
magnitude predicted in the simplest constructions.  There is
persistent difficulty in keeping Higgs doublets light against
radiative corrections, while giving large masses to their partners
under the GUT group, which carry $SU(3)\ll{\rm color}$ charges. These
scalar triplets thus interact with the quark and lepton sector via the
Yukawa couplings, causing unacceptably large proton decay and other
thus unobserved phenomena.  This issue, the doublet-triplet splitting
problem, is a major challenge for GUT model building.

In this Letter we explore an alternative framework that reproduces a
major success of GUTs, namely $U(1)\ll{em}$ charge quantization, while
at the same time eliminating the unwanted matter and massive gauge
fields that plague GUT models.

Our proposal begins by treating the SM group ${\cal G}\ll{\rm SM} =
SU(3)_{\rm{color}} \times SU(2) \times U(1)\ll Y$ as a local symmetry,
some part of which $H\subseteq {\cal G}\ll{\rm SM}$ is embedded in a
larger global group $G$.
\bbb
U(1)\ll Y \subseteq H \subseteq {\cal G}\ll{\rm SM}\ , \qquad H \subset G\ .
\eee
We work in a theory where $G$ is a nonlinearly realized symmetry. We
never consider it to be linearly realized and spontaneously broken,
nor gauged. In this work we will take $G = SU(2)_G$ and $H = U(1)_H$,
and comment only briefly on generalizations which will be explored in
a followup work.

Our first step is to consider the dynamics of a nonlinear $\s$-model
(NLSM) on the coset space $C \equiv G/H$, which has a natural set of
global symmetries $G$ realized as isometries of the coset $C$ derived
from its natural left $G$-action. With $G = SU(2)_G$ and $H = U(1)_H$,
we have $C = \icp 1$. Our second step is to add matter fields coupling
to the NLSM that are linear representations of $H$, yet have dynamics
that are invariant under $G$. Since $G$ is partially nonlinearly
realized, the construction of such actions is nontrivial, and
constitutes a major portion of the detail of this Letter. For the third
step, we gauge the linearly realized symmetry $H$, so that the matter
fields $\chi$ are incorporated into the Standard Model, and the
Nambu-Goldstone (NG) bosons described by the fields of $C$ are coupled
to it through gauge interactions as well.

We find that in order to write down kinetic terms for the matter that
are nonsingular and invariant under the full group $G$ everywhere on
$C$, the $U(1)_H$ charges of the matter fields $\chi$ must be
quantized in units of \it half \rm the charge of the Nambu-Goldstone
boson. Thus the structure of the NLSM itself and its interactions with
matter eliminates the need for a GUT group, with all its
phenomenologically troublesome baggage, in order to explain the
quantization of electric charge.

The organization of the Letter is as follows. In Section \ref{sec:cp1}
we introduce the $\icp 1$ model. We derive the couplings of matter
fields $\chi$ to $C$ under the assumption of exact $SU(2)_G$ symmetry,
and in particular show that the $U(1)\ll H$ charge of $\chi$ must be
an integer multiple of $\hh$ the charge of a NG boson. In Section
\ref{sec:pheno} we then gauge the $U(1)\ll H$ and examine the
phenomenological consequences of embedding this into the Standard
Model gauge group as $U(1)\ll Y$. We discuss some further aspects of
the model, including briefly commenting on generalizations, and
conclude in Section \ref{sec:conc}.

\section{The $\icp 1$ model $SU(2)_G / U(1)_H$}
\label{sec:cp1}

We start by introducing our conventions for $\icp 1$. With a
straightforward analysis, we can show charge quantization outside of
the usual setting (e.g.~a GUT or monopole). We derive a charge
quantization condition by considering a complex charged matter
field. This field will transform linearly under the unbroken $U(1)_H$,
but nonlinearly under the $SU(2)_G$. By explicitly determining the
transformation properties for the field, and requiring that all
transformations be smooth over the entire manifold, we find that the
charge of the field is quantized: the charge is a half-integer
multiple of the Nambu-Goldstone boson's charge.

If we identify the $U(1)_Y$ of the SM as $\icp 1$, we can match the
known hypercharges of the SM by fixing the NG boson charge. This can
explain charge quantization in the SM. Additionally, the NG boson is a
color-neutral, fractionally charged particle which can be a stable
component of dark matter or have applications in nuclear physics. We
discuss this briefly in the following section.

\subsection{Symmetries and dynamics of the NLSM}

The complex projective space \icp 1 has two (complex) homogeneous
coordinates, $\phi_{1,2}$. These satisfy $(\lambda\phi_1,
\lambda\phi_2) = (\phi_1, \phi_2)$. We can then define affine
coordinates as their ratio, namely $z_+ \equiv v\phi_1/\phi_2$, where
the vev, $v$, is used to give $z_+$ mass dimension one.

We label the infinitesimal generators of $SU(2)_G$ as $T\ll\pm$ and
$T\ll 0$. On the field $z\ll +$ the generators act
as
\begin{subequations}
\begin{align}
\d\ll{T\ll +}\aat z\ll + &= - {1\over v} \cc z\ll + \sqd \ ,\\
\d\ll{T\ll -}\aat z\ll + &= v\ ,\\
\d\ll{T\ll 0}\aat z\ll + &= + z\ll + \ ,\label{normA}
\end{align}
\end{subequations}
so we can write the action of $SU(2)_G$ on holomorphic functions of $z\ll +$ as
\begin{subequations}
\begin{align}
\d\ll{T\ll +}\equiv \d\ll{T\ll +}\hol &= -{1\over v} \cc z\ll + \sqd  \cc\pp\ll{z\ll +}\ ,\label{normBBB}\\
\d\ll{T\ll -}\equiv \d\ll{T\ll -}\hol &= v\cc \pp\ll{z\ll +}\ , \label{normBB}\\
\d\ll{T\ll 0}\equiv \d\ll{T\ll 0}\hol &= z\ll + \cc\pp\ll{z\ll +}\ ,\label{normB}
\end{align}
\end{subequations}
which obey the commutators
\begin{subequations}
\begin{align}
[\d\ll{T\ll 0}, \d\ll{T\ll \pm}] &= \pm \d\ll{T\ll\pm}\ , \label{normC}\\
[\d\ll{T\ll +}, \d\ll{T\ll -}] &= 2 \d\ll{T\ll 0} \label{normD}\ .
\end{align}
\end{subequations}
The action on antiholomorphic functions is obtained by the replacements
\begin{align}
{\rm holomorphic} &\to {\rm antiholomorphic}\ ,\nn \\
T\ll 0 \to - T\ll 0 \ ,  \qquad & \qquad \qquad T\ll{\pm} \to -T\ll{\mp}\ , \label{AntiHoloRules}
\end{align}
that is,
\begin{subequations}
\begin{align}
\d\ll{T\ll +}\ant &= -v\cc \pp\ll{\zb\ll +}\ ,\\
\d\ll{T\ll -}\ant &= +{1\over v} \cc \zb\ll + \sqd  \cc\pp\ll{\zb\ll +}\ ,\\
\d\ll{T\ll 0}\ant &= -\zb\ll + \cc\pp\ll{\zb\ll +}\ ,
\end{align}
\end{subequations}
when acting on antiholomorphic coordinates.  The full generators are simply a sum of
holomorphic and antiholomorphic pieces, 
\begin{align}
\d\ll{T\ll \pm} \equiv \d\hol\ll{T\ll\pm} + \d\ant\ll{T\ll\pm}\ , \hskip.3in  \d\ll{T\ll 0} \equiv \d\hol\ll{T\ll 0} + \d\ant\ll{T\ll 0}\ .
\label{FullGens}\end{align}
We will omit the labels (hol.) and (ant.) when clear.

The same $SU(2)\ll G$-action on holomorphic functions, eq.~\rr{normBBB}-\rr{normB}, can be
written, using the chain rule, in terms of the variable $z\ll - \equiv
v\sqd / z\ll +$:
\begin{subequations}
\begin{align}
\d\ll{T\ll +} &= + v\cc \pp\ll{z\ll -}\ ,\\
\d\ll{T\ll -} &= - {1\over v} \cc z\ll - \sqd  \cc\pp\ll{z\ll -}\ ,\\
\d\ll{T\ll 0} &= - z\ll - \cc\pp\ll{z\ll -}\ ,
\end{align}
\end{subequations}
which obeys the same commutators, as it must.

The nonlinearly realized $SU(2)_G$ symmetry fixes the form of the
kinetic term for the Goldstone bosons uniquely, up to an overall
coefficient.  The metric is K\"ahler with respect to the same complex
coordinate, $z_+$, with which the symmetry is holomorphic, and the
K\"ahler potential is fixed to be proportional to $v\sqd\cc {\rm
  ln}(v\sqd + |z\ll +|\sqd)$.

We are working so far purely in the target space \icp 1, which is a
complex K\"ahler manifold, and the action of $SU(2)\ll G$ on the
holomorphic fields $z\ll +$ is holomorphic.  The K\"ahler property of
the metric and the holomorphy of the isometries are automatic
consequences of the geometry of the $\icp 1$ with Fubini-Study metric.
The model therefore admits a natural $\mathcal{N} = 1$
supersymmetrization, and we analyze principally the supersymmetric
version in this Letter. However, our conclusion that charge is
quantized for matter coupled to the sigma model holds even if
supersymmetry is broken explicitly. We will comment further on this
point in Section \ref{subsec:kinetic}. The phenomenology of the model,
with supersymmetry, is discussed in Section \ref{sec:pheno}.

\subsection{Consistency conditions for matter}
Now let us introduce a complex charged field $\chi,$
that transforms linearly under $T\ll 0$,
\begin{equation}
\d\ll{T\ll 0} = \a\cc\chi\cc \pp\ll \chi\ ,
\end{equation}
with $\alpha$ the $U(1)_H$ charge, and in some not yet determined
nonlinear way under $T\ll{\pm}$.

We have:
\begin{subequations}
\begin{align}
\d\ll{T\ll +} &= F\ll +(\chi, z\ll +)\pp\ll\chi \ ,\\
\d\ll{T\ll -} &= F\ll -(\chi, z\ll +)\pp\ll\chi \ppo
\end{align}
\end{subequations}
For now we only define the action of the generators in the southern
hemisphere ($z\ll - \neq 0$).  We will examine the extension to the
northern hemisphere ($z\ll + \neq 0$) later; we shall see that the
condition that the action extends smoothly will impose precisely the
condition of charge quantization.

Now let us examine the conditions for the closure of the commutation relations. The full generators are:
\begin{subequations}
\begin{align}
\d\ll{T\ll +} &= F\ll + (\chi, z\ll +)\pp\ll\chi - {1\over v} \cc z\ll + \sqd \cc\pp\ll{z\ll +}\\
\d\ll{T\ll -} &= F\ll - (\chi, z\ll +)\pp\ll\chi + v\cc \pp\ll{z\ll +}\\
\d\ll{T\ll 0} &= \a\cc\chi\cc \pp\ll \chi + z\ll + \cc\pp\ll{z\ll +} \ .
\end{align}
\end{subequations}

The commutation relations become first-order differential equations on $F\ll\pm$:
\begin{subequations}
\begin{align}
 [\d\ll{T\ll 0} , \d\ll{T\ll +}] &= 
\left(-\a\cc F\ll + + \a\chi F\ll{+,\chi} + z\ll + \cc F\ll{+,z\uu +}\right) \pp\ll\chi
\nonumber\\ & \qquad %
 - {1\over v}z\ll + \sqd\pp\ll{z\ll +}\\
[\d\ll{T\ll 0} , \d\ll{T\ll -}] &= 
\left(-\a\cc F\ll - +  \a\chi F\ll{-,\chi} + z\ll + F\ll{-,z\uu +}\right) \pp\ll\chi
\nonumber\\ & \qquad %
 - v \cc  \pp\ll{z\ll +}\\
[\d\ll{T\ll +} , \d\ll{T\ll -}] &=
(F\ll + \cdot F\ll {-,\chi} - {1\over v} z\ll + \sqd F\ll{-,z\ll +} - F\ll - \cdot F\ll {+,\chi}\nonumber\\
&\qquad - v F\ll{+,z\ll +}) \pp\ll\chi + 2 z\ll + \pp\ll{z\ll +}
\end{align}
\end{subequations}
and by matching with the algebra and the generators above we have
\begin{subequations}
\begin{align}
\hskip-.3in
-\a F\ll + + \a\chi F\ll{+,\chi} + z\ll + F\ll{+,z\uu +}  
&= + F\ll +\\
\hskip-.3in
-\a F\ll - +  \a\chi F\ll{-,\chi} + z\ll + F\ll{-,z\uu +}
&= - F\ll -\\
\hskip-.5in F\ll + \cdot F\ll {-,\chi}
 - {1\over v} z\ll + \sqd F\ll{-,z\ll +}
- F\ll - \cdot F\ll {+,\chi} - v F\ll{+,z\ll +}
&= + 2 \a\cc\chi\ .
\end{align}
\end{subequations}

It's easiest to begin by solving the first two equations, which
are linear first-order PDEs.  The general solutions to the first
and second equations are
\begin{align}
F\ll + &= v\uu{-\a}\cc z\ll +\uu{\a + 1} \cc f\ll +\left(
 {{v\uu{\a - 1} \cc \chi }\over{ z\ll +\uu\a}} \right) \kko\\
F\ll - &= v\uu{2-\a}\cc z\ll +\uu{\a - 1} \cc f\ll -\left(
 {{v\uu{\a - 1} \cc \chi }\over{ z\ll +\uu\a}} \right) \kko
\end{align}
where $f\ll\pm$ are two arbitrary functions of the dimensionless
ratio $(v\uu{\a - 1} \cc \chi) / z\ll +\uu\a$.

Now restrict to transformations linear in $\chi$ (but involving \it a
priori \rm unknown nonlinear functions of $z\ll +$.)  We do not lose
any generality by doing this.  If $F\ll\pm$ contain $\chi\uu 0$ terms,
we can remove them by redefining $\chi$ additively by a function of
$z\ll +$.  Then $F\ll\pm$ can be assumed to be linear in $\chi$ plus
terms of order $\chi\sqd$ and higher.  By examining the action on
$\chi$ at small $\chi$, we can see that the transformations must close
among themselves at the order $\chi\uu 1$ level.  So we lose no
generality by taking the transformations of $\chi$ to be strictly
linear in $\chi$ (but involving unknown nonlinear functions of $z\ll
+$): The constraints on the order $\chi\uu 1$ terms in the
transformation law for $\chi$ are independent of the order $\chi\sqd$
and higher-order terms.

We then take the transformations to be 
smooth at the south pole $z\ll + = 0$, which fixes
\begin{align}
F\ll - &= 0\ ,\\
F\ll + &= ({\rm const.}) \cc z\ll + \cc\chi\ .
\end{align}

Imposing the third commutator equation fixes the constant, and we find
\begin{align}
F\ll - &= 0\ ,\\
F\ll + &= - {{2\a}\over v} \cc z\ll + \cc\chi\ .
\end{align}

So all in all, the holomorphic generators are
\begin{subequations}
\begin{align}
\d\hol\ll{T\ll +} &=  - {{2\a}\over v} \cc z\ll + \cc\chi\pp\ll\chi - {1\over v}
\cc z\ll +\sqd \pp\ll{z\ll +}\kko\label{MatterTransA}\\
\d\hol\ll{T\ll -} &=  v\pp\ll{z\ll +}\kko\label{MatterTransB}\\
\d\hol\ll{T\ll 0} &= \a\cc \chi\pp\ll\chi + z\ll + \pp\ll{z\ll +}\ppo\label{MatterTrans}
\end{align}
\end{subequations}
Again, the action on the antiholomorphic $\chi\dag$ is obtained by
extending the rules in eq.~\rr{AntiHoloRules} to include $\chi\dag$:
\begin{subequations}
\begin{align}
\d\ant\ll{T\ll +} &= -v\cc \pp\ll{\zb\ll +}\ ,\\
\d\ant\ll{T\ll -} &= + {{2\a}\over v} \cc \zb\ll + \cc \chi\dag\pp\ll{\chi\dag} +{1\over v} \cc \zb\ll + \sqd  \cc\pp\ll{\zb\ll +}\ ,\\
\d\ant\ll{T\ll 0} &= - \a\chi\pp\ll{\chi\dag} -\zb\ll + \cc\pp\ll{\zb\ll +}\ ,
\end{align}
\end{subequations}

\subsection{Proof of charge quantization}

Now we would like this to be well-defined in the northern hemisphere,
which is to say, manifestly smooth in coordinates $z\ll - = v\sqd /
z\ll +$.  In order to make this so, we must also transform the field
$\chi$ as well, to a new field $\chi\pr$.  Since the $SU(2)_G$
transformations are linear in $\chi$, the transformation between
$\chi$ and $\chi\pr$ should be linear in $\chi$.  That is, $\chi$
should transform as a section of a vector bundle over $\icp 1$.  A
general linear change of basis is:
\begin{equation}
\chi\pr \equiv f(z\ll +) \cc \chi\kko\llsk \chi = {1\over{f(z\ll +)}} \cc \chi\pr\kko
\end{equation}
and we would like to choose $f(z\ll +)$ such that the $SU(2)_G$
transformations act smoothly in the northern hemisphere as well as in
the southern hemisphere.

In northern-hemisphere coordinates $z\ll - , \chi\pr$, we have
\begin{subequations}
\begin{align}
\pp\ll\chi &= {{\pp \chi\pr}\over{\pp\chi}} \cc\pp\ll{\chi\pr} 
= f(z\ll +)\cc\pp\ll{\chi\pr}
\kko\llsk\\
\pp\ll{z\ll +} &= - v\uu{-2} \cc z\ll - \sqd \pp\ll{z\ll -} + f\pr(z\ll +) 
\chi \pp\ll{\chi\pr}\nonumber\\
& = - v\uu{-2} \cc
z\ll - \sqd \pp\ll{z\ll -} + {{f\pr(z\ll +)}
\over{f(z\ll +)}} \chi\pr \pp\ll{\chi\pr}
\end{align}
\end{subequations}

So
\begin{subequations}
\begin{align}
\chi\pp\ll\chi &= \chi\pr\pp\ll{\chi\pr}\kko\\
z\ll + \cc \chi\pp\ll\chi &= z\ll + \cc \chi\pr\pp\ll{\chi\pr}\kko\\
z\ll + \pp\ll{z\ll + } &= - z\ll - \pp\ll{z\ll -} 
+ {{z\ll + \cc f\pr(z\ll +)}\over{f(z\ll +)}}\cc\chi\pr\pp\ll{\chi\pr}
\kko\\
z\ll + \sqd \pp\ll{z\ll +} &= - v\sqd\cc \pp\ll{z\ll -}
+ {{z\ll +\sqd
 \cc f\pr(z\ll +)}\over{f(z\ll +)}}\cc\chi\pr\pp\ll{\chi\pr}
\end{align}
\end{subequations}
The unbroken $U(1)\ll H$ at the north pole is the same as the unbroken
$U(1)\ll H$ at the south pole, because any two antipodal points are
fixed by the same rotation generator.  Therefore the matter $\chi$ at
the south pole and $\chi\pr$ at the north pole must each have definite
eigenvalues under the same $U(1)\ll H$.  Thus we can take take the
same generator $T\ll 0$ to act simultaneously on $\chi$ and
$\chi^\prime$ as multiplication by constants. It follows that the
transition function $f(z\ll +)$ is a pure power, $f(z\ll +) = K\ll f
\cc z\ll +\uu p = v^{2p} K\ll f \cc z\ll -\uu {-p}$, where $K\ll f$ is
an arbitrary constant that can be absorbed into the normalization of
$\chi\pr$.  Then we have:
\begin{subequations}
\begin{align}
\chi\pp\ll\chi &= \chi\pr\pp\ll{\chi\pr}\kko\\
z\ll + \cc \chi\pp\ll\chi &= z\ll + \cc \chi\pr\pp\ll{\chi\pr}
= v\sqd\cc z\ll -\uu{-1} \cc \chi\pr\pp\ll{\chi\pr}\kko\\
z\ll + \pp\ll{z\ll + } &= - z\ll - \pp\ll{z\ll -} 
+ {{z\ll + \cc f\pr(z\ll +)}\over{f(z\ll +)}}\cc\chi\pr\pp\ll{\chi\pr}\nonumber\\
&=  - z\ll - \pp\ll{z\ll -}  + p \cc\chi\pr\pp\ll{\chi\pr}\kko\\
z\ll + \sqd \pp\ll{z\ll +} &= - v\sqd\cc \pp\ll{z\ll -} + p\cc z\ll + \cc\chi\pr\pp\ll{\chi\pr}\nonumber\\
&= - v\sqd\cc \pp\ll{z\ll -} + v\sqd\cc p\cc z\ll -\uu{-1} \cc\chi\pr\pp\ll{\chi\pr} \kko
\end{align}
\end{subequations}

Then form of the transformations $\d\ll{T\ll 0}, \d\ll{T\ll \pm}$ in
the northern hemisphere is
\begin{subequations}
\begin{align}
\d\ll{T\ll 0} &= - z\ll - \pp\ll{z\ll -} + \left(\alpha + p\right) \chi\pr\pp\ll{\chi\pr}\kko\\
\d\ll{T\ll -} &= -{z\ll -\over v} \left ( z\ll -\cc \pp\ll{z\ll -} - p \chi\pr\pp\ll{\chi\pr}\right ) \kko \\
\d\ll{T\ll +} &=  v\cc \pp\ll{z\ll -} - v\left(p+ 2\a\right) z\ll -\uu{-1} \chi\pr\pp\ll{\chi\pr} \ppo
\end{align}
\end{subequations}

We want this transformation to be nonsingular everywhere in the
northern hemisphere $z\ll - \in\IC$, including the north pole, where
$z\ll - = 0$.  This imposes the condition
\begin{equation}
p = - 2\a\ .
\end{equation}

The change of variables between $\chi$ and $\chi\pr$ must be
single-valued everywhere in the overlap region $z\ll\pm\in\IC -
\{0\}$, which fixes
\begin{equation}
p\in\mathbb{Z}\ .
\end{equation}
This forces $\a$ to live in $\mathbb{Z} / 2$, giving the
charge-quantization condition.  

In mathematical terms, the matter field $\chi$ can be thought of as
the fiber coordinate of a line bundle over $\IC\IP(1)$, and the charge
of the matter field corresponds to half the degree of the line bundle.
The charge quantization for matter coupled to $\IC\IP(1)$ follows from
the Birkhoff-Grothendieck theorem \cite{cp1b,*cp1g},
which classifies holomorphic bundles over $\IC\IP(1)$.

\subsection{Kinetic terms for matter fields}\label{subsec:kinetic}

For a more concrete way of understanding the origin of charge
quantization, we can think of the condition on the charge as a
requirement on the kinetic terms for the fields of the coupled
matter-NG system.
The kinetic terms must be smooth in both hemispheres
and consistent under coordinate changes from hemisphere to hemisphere.
We take the K\"ahler potential to be quadratic in $\chi, \chi\dag$ and
invariant under the separate flavor symmetry that acts as a phase
rotation on $\chi$.  Imposing invariance under the combined generators $\d \equiv \d\hol + \d\ant$
forces the K\"ahler potential for the matter to be
\begin{equation}
K\ll{\rm matter} = \left(1 + {{|z\ll +|\sqd}\over{v\sqd}}\right)\uu{- 2\alpha} |\chi|\sqd\ ,
\end{equation}
in the southern hemisphere, up to an overall coefficient of
proportionality that can be absorbed into the normalization of $\chi$.
In the northern hemisphere, if we demand that the kinetic term have
the same form under the simultaneous replacement $(z\ll +, \chi) \to
(z\ll -, \chi\pr)$, then equality of the northern and southern
hemisphere expressions uniquely fixes $\chi\pr =
{{v\uu{2\a}}\over{z\ll +\uu{2\a}}}\cc\chi$.  This transformation is
single valued if and only if $p \equiv -2\a \in \mathbb{Z}$.

We have so far assumed that our theory has ${\cal N} = 1$
supersymmetry, which restricts the the combined metric defining the
kinetic term for the NG bosons $z\ll +$ and matter fields $\chi$, to
be K\"ahler.  This assumption simplifies our derivations considerably.
Unlike the case for $\icp 1$ in isolation, the K\"ahler property of
the combined kinetic term for $z\ll +$ and $\chi$, is not an automatic
consequence of the (bosonic) global symmetries.  In the absence of
SUSY, the K\"ahler condition on the kinetic term is not preserved
under renormalization group (RG) flow. If we were to add mass terms for
the fermions, breaking supersymmetry, this would alter the quantum
properties of the theory and the quantum-corrected kinetic term for NG
bosons and matter fields need not in general be a K\"ahler
metric. However, charge quantization is a RG-invariant property, and
therefore our demonstration of charge quantization for matter coupled
to the NLSM does not depend on supersymmetry or on the existence of a
K\"ahler metric on field space at all.

\section{Phenomenology of the \icp 1 Model}
\label{sec:pheno}

To show that we can use this $\icp 1$ model as the $U(1)_Y$ part of
the SM, we must be able to reproduce all of the hypercharges. There
is an inherent freedom in the choice of the charge of the NG
boson, and all other charges are proportional to this
one.\footnote{While the hypercharges are easy to reproduce in this
  model, in other models the charges will be related in a more
  complicated manner.} We will fix this charge through a type of
``minimality:'' the charge of the NG boson is chosen such that the
smallest possible hypercharge is $1/6$, the smallest
charge in the SM.

The generator of weak hypercharge is then defined as one-third the
Cartan generator $T\ll 0$, normalized as in eq.~\rr{MatterTrans}:
\begin{equation}
Q_Y = \frac{1}{3} T\ll 0.
\end{equation}
We have $T\ll 0= 1$ for the Nambu-Goldstone boson, so it has
hypercharge $1/3$. All other matter fields then have hypercharge
\begin{equation}
Q_Y = \frac{n}{6},
\end{equation}
where $n/2$ is the charge in units of the charge of the NG boson. Thus
we see that we can easily reproduce the SM hypercharges with matter
coupled to \icp 1.

It is clear that at this stage we have a massless, fractionally
charged, and stable Nambu-Goldstone boson. The stability is ensured by
its fractional charge under $U(1)\ll{\rm em}$: The NG boson is a color
singlet, weak isosinglet, and has $U(1)\ll{\rm em}$ electric charge
equal to its weak hypercharge, $1/3$.  It is therefore absolutely
stable as there are no fractionally charged color singlets to which it
can decay.  Also, since it carries electric charge, once we gauge the
$U(1)_H$ gauge interactions will give it a mass, of order
$\sqrt{\alpha} M_{G/H}$. We can take\footnote{We note here that this
  model can produce cosmic strings (but \textit{not} magnetic
  monopoles), but at a scale much above the inflation scale, so they
  are of no concern.} $M_{G/H} \sim M_{\rm Planck}$, and thus the NG
boson is very heavy. However, with supersymmetry we expect the mass to
be protected down to about $\sqrt{\alpha} M_{\rm Susy}$.

Gauging the $U(1)_H$ explicitly breaks the $SU(2)_G$. However, this is
controlled by the gauge coupling; as we take the coupling to zero, we
flow continuously to the exact global case. Charge quantization cannot
be changed by this explicit breaking. In the followup paper we will
consider more general explicit breaking and its relation to the charge
quantization condition.

In a supersymmetric theory, the NG boson is accompanied by a fermion
partner and hence we have gauge anomalies. To cancel the anomalies we
can introduce a chiral matter multiplet whose $U(1)_H$ charge is
conjugate to that of the NG multiplet. This matter multiplet can also
cancel the nonlinear sigma model anomalies
\cite{Moore:1984dc,*diVecchia:1984jh,*Cohen:1984js,*Yanagida:1985jc}. It
may be natural that the fermions receive a Dirac mass of order the
gravitino mass once supersymmetry and the R-symmetry are broken. These
Dirac fermions can be lighter than the NG boson, depending on the
supersymmetry breaking mechanism. Here, however, we will assume that
the NG boson is the lightest among them. In this case it could be
probed by the LHC when the supersymmetry scale is not too high, say of
order $\sim 10$ TeV.

This NG boson is then a candidate for dark matter. Charged dark matter
was first considered and constrained tightly some time ago
\cite{Goldberg:1986nk,*DeRujula:1989fe,*Basdevant:1989fh,*Dimopoulos:1989hk,Gould:1989gw},
and has more recently been revived
\cite{Chuzhoy:2008zy,*McDermott:2010pa}. The immediate concern is for
production of this charged particle in the early universe, which is
stringently constrained both astrophysically and terrestrially. One
can simply take the reheating temperature to be far enough below the
mass of the Nambu-Goldstone so that production is negligible. Even for
low-scale supersymmetry, this is plausible. With some small amount of
production to avoid constraints, however, the NG boson could be a
component of dark matter.

We see that it is possible then to have a light, fractionally-charged
boson that is accessible in colliders while satisfying current experimental
constraints. Besides the discovery of such a particle to be a rather
unique prediction of this scenario (given the assumptions of the
relevant scales), what are some other consequences of such an unusual
fundamental particle?

This Nambu-Goldstone boson could have profound consequences in nuclear
physics. One very enticing idea is to use such a particle to catalyze
nuclear fusion \cite{fusion1,*fusion2,*fusion3,*fusion4}. By forming a
bound state between the NG boson and, for instance, the deuteron, the
Coulomb potential is screened. This lowers the required energy (or
temperature) for fusion to occur. While a distant thought, this could
make fusion-based energy accessible. As a stable, charged, and heavy
particle, the NG boson could also be used as a probe of the structure
of heavy nuclei by analyzing the interactions as it penetrates into
the nucleon.

While we see that it is possible to have the NG boson as a light state
in the low energy theory, this is not an unavoidable prediction. It is
entirely possible that there are no residual effects besides charge
quantization. This differs greatly from the standard GUT scenario and
we will not observe or predict gauge coupling unification. Even with
no new phenomena to be seen (e.g.~if the supersymmetry scale is very
high) we still have charge quantization in the SM via the \icp 1
structure.

\section{Discussion and Conclusions}
\label{sec:conc}

In this Letter we have only considered the \icp 1 model, but it is
clear there should be a generalization to other cases. The \icp 1
model already encapsulates all of the important aspects of our
approach and has clear phenomenological applications. Furthermore, by
analyzing this model, we already learn about more complicated
constructions.

Consider the more general model \icp k. Rather than directly following
the procedure of Section \ref{sec:cp1}, we can instead add mass terms
to the additional Nambu-Goldstone modes and flow down to the \icp 1
model we have already studied. It is straightforward to do this and
arrive at a generalization of the quantization formula we have derived
in this Letter. In a followup paper we work this out in detail, along
with other models, and consider their phenomenological consequences.

To summarize, we have found a new way of quantizing electromagnetic
charge, without introducing the problems typically associated with
standard GUTs. The basic idea is simple: consider a nonlinear sigma
model, where there is an unbroken $U(1)$ factor. From the constraints
of the NLSM, namely that matter fields and their transformations under
$G$ are well-defined everywhere, the charges of any matter are
restricted and related to the NG boson charge. We explicitly analyzed
the \icp 1 model $SU(2)_G/U(1)_H$, where we see all of the matter
charges are half-integer multiples of the NG boson charge. This
$U(1)_H$ is gauged and identified as the weak hypercharge group of the
SM. Thus, the NG boson has electromagnetic charge $1/3$.

This model does not have any of the usual difficulties associated with
GUTs: there are no extra gauge bosons to mediate proton decay, no
monopoles, no colored Higgs partners, and so on. Charge quantization,
a major success of GUTs, is a consequence of the geometry and
dynamics of the nonlinear sigma model. A product of this construction
is a Nambu-Goldstone boson that can appear in the low energy
theory. The NG boson is not forced to be very light (for instance,
even if protected by supersymmetry, the breaking scale could be very
high), so it is possible to have no new low energy effects but still
have successful charge quantization.

Given the right circumstances, though, this NG boson can play a very
interesting phenomenological role. It is absolutely stable, has
fraction electromagnetic charge, and can even be light enough to be
seen in present colliders. Such a charged particle is not a typical
prediction of supersymmetric or many other models. While avoiding
cosmological constraints it can have a profound impact on nuclear
physics, especially fusion. In an optimistic scenario then, we can see
a smoking gun signature of charge quantization which can have very
promising everyday applications.

\vspace{2pc}

\begin{acknowledgments}
\noindent
T.T.Y.~would like to thank Yuji Tachikawa for discussions about \icp
1. J.K.~would like to thank Brian Feldstein for conversations about
charged dark matter. This work was supported by the World Premier
International Research Center Initiative (WPI Initiative), MEXT,
Japan.  The work of S. H. was also supported in part
by a Grant-in-Aid for Scientific Research (22740153) from the Japan Society for Promotion of Science (JSPS).
\end{acknowledgments}

\renewcommand{\bibsection}

{\section*{References}}
\bibliography{cp1-cq-refs}

\end{document}